\documentclass[twocolumn]{aastex62}

\accepted{for publication in the Astrophysical Journal Letters}

\shorttitle{Asymmetric Metallicity Distribution}
\shortauthors{An}

\begin{document}

\title{Asymmetric Mean Metallicity Distribution of the Milky Way's Disk}

\author{Deokkeun An}
\affiliation{Department of Science Education, Ewha Womans University, 52 Ewhayeodae-gil, Seodaemun-gu, Seoul 03760, Republic of Korea; deokkeun@ewha.ac.kr}

\begin{abstract}

I present the mean metallicity distribution of stars in the Milky Way Galaxy based on photometry from the Sloan Digital Sky Survey. I utilize an empirically calibrated set of stellar isochrones developed in previous work to estimate the metallicities of individual stars to a precision of $0.2$~dex for reasonably bright stars across the survey area. I also obtain more precise metallicity estimates using priors from the {\it Gaia} parallaxes for relatively nearby stars. Close to the Galactic mid-plane ($|Z|<2$~kpc), a mean metallicity map reveals deviations from the mirror symmetry between the northern and southern hemispheres, displaying wave-like oscillations. The observed metallicity asymmetry structure is almost parallel to the Galactic mid-plane, and coincides with the previously known asymmetry in the stellar number density distribution. This result reinforces the previous notion of the plane-parallel vertical waves propagating through the disk, in which a local metallicity perturbation from the mean vertical metallicity gradient is induced by the phase-space wrapping of stars in the $Z$-$V_Z$ plane. The maximum amplitude of the metallicity asymmetry ($\Delta$[Fe/H]$\sim0.05$) implies that these stars have been pulled away from the Galactic mid-plane by an order of $\Delta|Z|\sim80$~pc as a massive halo substructure such as the Sagittarius dwarf galaxy plunged through the Milky Way. This work provides evidence that the {\it Gaia} phase-space spiral may continue out to $|Z|\sim1.5$~kpc.

\end{abstract}


\section{Introduction}

The Galactic disk's deviation from the mirror symmetry with respect to its mid-plane was first discovered by \citet{widrow:12} from star counts based on photometry in the Sloan Digital Sky Survey (SDSS), and was then characterized in depth by \citet{yanny:13}. This density asymmetry with a maximum amplitude of $\sim10\%$ was recently confirmed by \citet{bennett:19} using {\it Gaia} parallaxes \citep{gaiadr2}. Other studies also found rich structures of vertical velocity distributions  in the local disk \citep{widrow:12,carlin:13,williams:13,carrillo:18,gaiadr2_kinematics,schonrich:18,bennett:19}, supporting the idea that the stellar disk is not in equilibrium in the direction perpendicular to the Galactic mid-plane.

The observed density asymmetry is almost parallel to the Galactic plane \citep{widrow:12}, and shows wave-like features with an excess/deficit of stars at $|Z|\approx0.4$, $0.9$, and $1.5$~kpc in the southern/northern Galactic hemisphere according to the most recent analysis \citep{bennett:19}. The Jeans length in the Galactic disk is about $2$~kpc \citep[e.g.,][]{widrow:12}. Therefore, if these are likely vertical waves propagating through the disk, they are stable and against gravitational collapse. The odd parity of the vertical density distribution suggests that the vertical disturbance is not caused by internal perturbations inside the Galactic disk \citep[but see][]{faure:14}. The most likely cause is excitation by external perturbations, such as the passage of the Sagittarius dwarf galaxy or interactions with halo substructures \citep[e.g.,][]{gomez:13,widrow:14}, which may be the same dynamical origin for other phase-space disturbances and/or corrugations found in the disk \citep[e.g.,][]{xu:15,antoja:18}.

Numerical simulations show that the vertical waves can survive for many hundreds of million years, until they disappear by phase mixing and resonant interaction \citep{weinberg:91,widrow:12,gomez:13,widrow:14}. If the density asymmetry is a manifestation of the oscillatory phenomenon, it should leave a distinctive signature in the vertical metallicity structure of the disk, as the Galactic disk has a steep metallicity gradient in the vertical direction \citep[$\sim0.3$~dex~kpc$^{-1}$ at the solar circle;][see references therein]{hayden:14,schlesinger:14} rather than the radial and azimuthal directions. Under a simplified assumption of pressure-supported waves, any displacement of stars is accompanied by a small change in the vertical metallicity gradient. Spectroscopic surveys may not be capable of detecting such differences in metallicity, because of a relatively poor completeness and a strong bias in the data.

In this Letter, I present a new metallicity map based on the SDSS imaging data, with an emphasis on the vertical metallicity structure. The photometric data is less susceptible to a sample bias, and covers a wide area of sky to a sufficient depth. A broadband photometric system, such as $ugriz$ in SDSS, can also be used to constrain metallicities of individual stars \citep[e.g.,][]{ivezic:08,an:09a,an:13,an:15,gu:15,ibata:17}. Although these photometric metallicity estimates have lower precision than spectroscopic determinations, photometry can be used to probe the chemical space for a significantly larger fraction of stars in the Galaxy. I summarize a general method of a photometric metallicity estimator in \S~2, which is followed by main results in \S~3. More details on the photometric metallicity distributions will be presented elsewhere (D. An et al.\ 2019, in preparation).

\section{Method}

I obtained the $ugriz$ photometric data from the Fourteenth Data Release (DR14) of the SDSS~IV \citep{sdss}\footnote{https://www.sdss.org/dr14/} that is based on the ``hyper-calibration'' procedure \citep{finkbeiner:16}. General details of the photometric metallicity estimator can be found in \citet{an:13}. In short, I constrained a metallicity,\footnote{[Fe/H] is used throughout this Letter to represent a metallicity of a star. Because the stellar isochrones employed in this work assume a certain relation between [Fe/H] and [$\alpha$/Fe], [Fe/H] increases with the bulk metallicity of a star ([M/H]), but they are not exactly same with each other for metal-poor stars with an enhanced $\alpha$ abundance.} distance, and mass (or effective temperature) for each star based on $ugriz$ photometry by searching for a minimum $\chi^2$ in a grid of stellar isochrones \citep{an:09b,an:13}. I took foreground extinctions from \citet{schlegel:98}, but with extinction coefficients at $R_V=3.1$ from \citet{schlafly:11}.

The stellar models include empirical color corrections to match the observed main sequences of several well-studied clusters. Since the publication of \citet{an:13}, a minor improvement in the fitting procedure (including a proper error estimate) was made and incorporated into the current study. The typical size of a statistical metallicity error is $0.2$~dex (and $\sim10\%$ error in distance) for metal-rich stars ([Fe/H]$\ga-1$), and is about $0.3$~dex for metal-poor stars with a reasonably accurate photometry ($0.03$~mag error in $u$, $0.01$~mag error in $gri$, and $0.02$~mag error in $z$). This approach assumes that all point-like sources detected in the survey are main-sequence stars; giants and dwarfs are not separable from the SDSS photometry alone. Although unrecognized giants can have a systematically higher photometric metallicity \citep[see][]{an:13}, a relative comparison between the northern and southern Galactic hemispheres should be less affected by these photometric contaminants. The same is true for unresolved binaries, of which photometric metallicity is overestimated.

The following criteria were used to select sample stars for the following analysis:
\begin{itemize}
\item Detected in all passbands
\item $E(B\, -\, V) < 0.10$
\item $|b| > 30\arcdeg$
\item $u < 20$
\item $4.5 < M_r < 7.5$
\item $\chi^2 < 5$
\item $\sigma_{\rm [Fe/H]} < 1.5$
\end{itemize}

My initial inspection of the photometric metallicity maps revealed regions with anomalously lower or higher metallicities than adjacent areas ($\Delta{\rm [Fe/H]}\sim0.5$). One of these cases is a strip centered at $l\approx28\arcdeg$ with a width of $\sim5\arcdeg$ that extends from the low latitude limit of the survey to the north Galactic pole. Because the strip is parallel to the scan direction of the SDSS imaging survey, I suspect that the offset is due to problems in the photometric calibration. The tendency of the metallicity difference is to make stellar colors redder in the northern Galactic hemisphere than in the southern hemisphere, which may explain a part of the color offsets between the two hemispheres found by \citet{schlafly:11}. For this reason, I did not include photometry in the boxed region ($\approx3\arcdeg\times90\arcdeg$) along a meridian at $l\approx28\arcdeg$. I also excluded photometry from a strip at $l=330\arcdeg$ with a width of $\sim5\arcdeg$ as well as a triangular patch surrounded by $45\arcdeg \la l \la 90\arcdeg$ and $30\arcdeg\la b\la45\arcdeg$  for the same reason.

Photometric solutions can be improved by setting a prior on an individual star's parallax ($\pi$). I chose a $1\arcsec$ search radius to cross-match sources with the Second Data Release (DR2) of the {\it Gaia} mission \citep{gaiadr2} and imposed an upper limit on parallax errors $\sigma_\pi/\pi<0.2$. I corrected parallaxes for the global parallax zero-point offset ($0.029$~mas) as suggested by the {\it Gaia} team \citep{lindegren:18}. Distance estimates from the full photometric solutions are within $5\%$ of the {\it Gaia} parallaxes over a wide range of metallicities.

\section{Results}

\begin{figure*}
\centering
\includegraphics[scale=0.6]{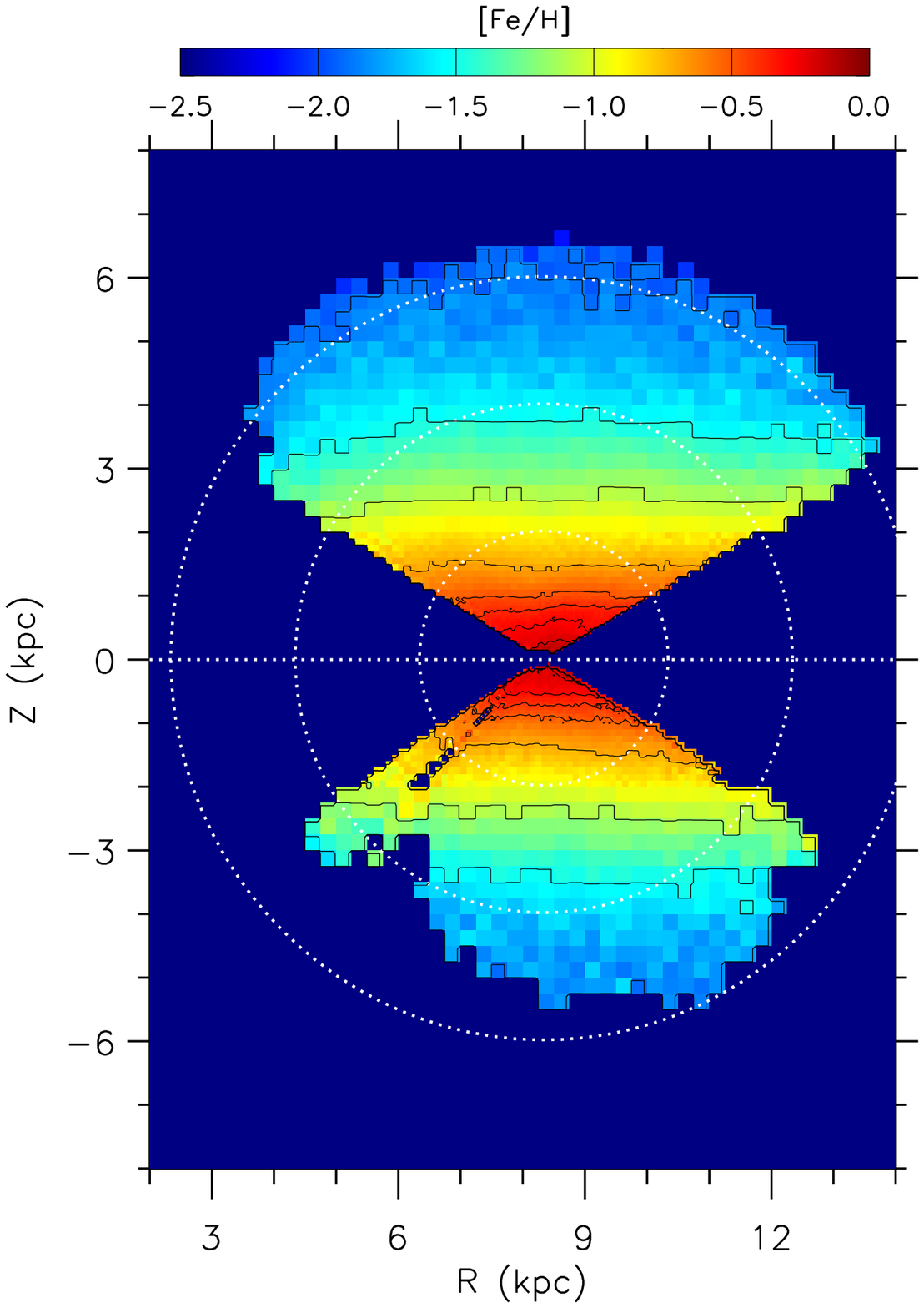}
\includegraphics[scale=0.6]{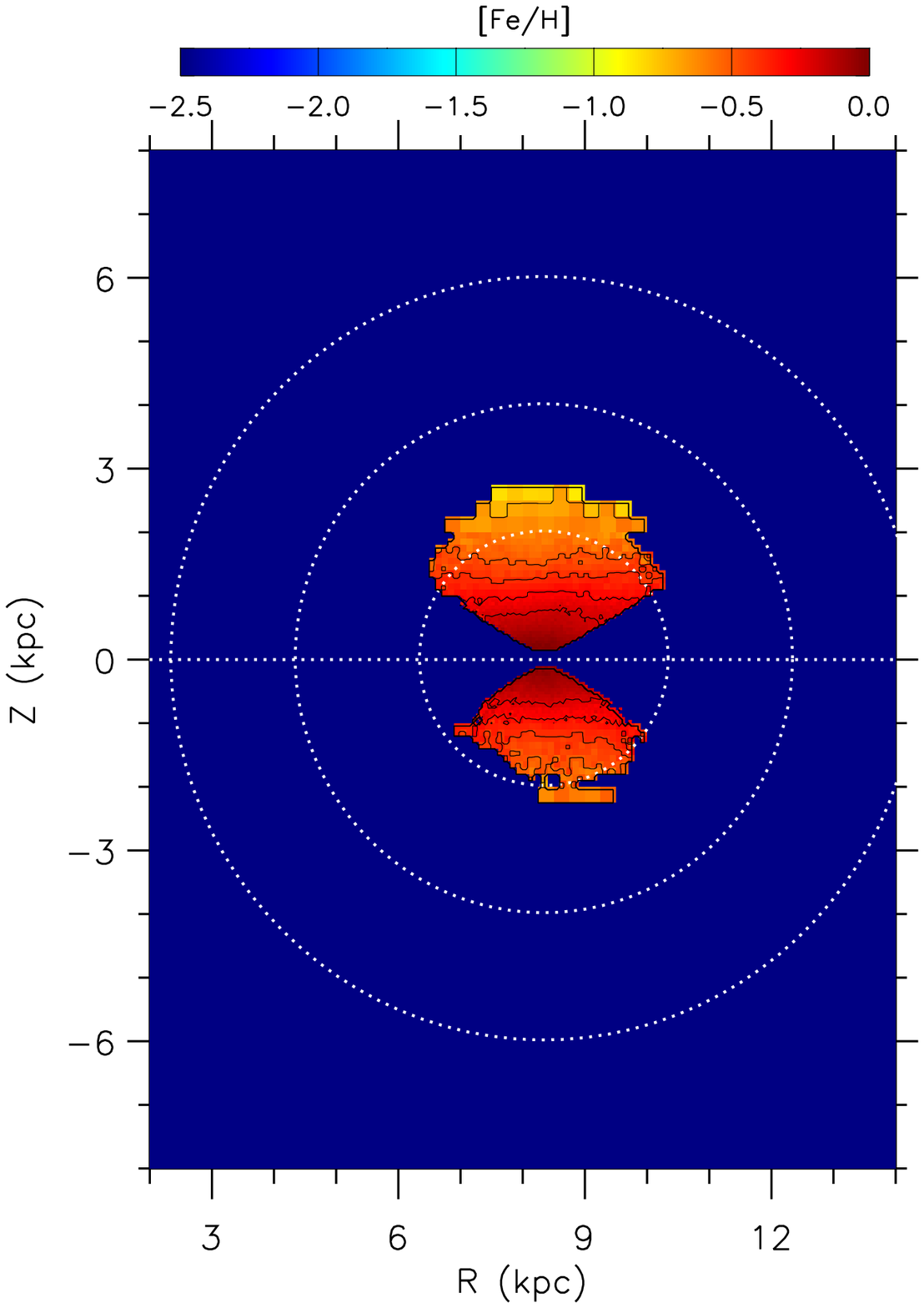}
\caption{Left panel: edge-on view of the metallicity distribution of stars ($|b|>30\arcdeg$) in the Galaxy from SDSS photometry in the Galactocentric cylindrical coordinate system. Mean metallicities from a generalized MDF are shown in each pixel of this 2D histogram, which has a dimension ($\Delta R$, $\Delta Z$) of ($0.05$~pc, $0.05$~pc), ($0.10$~pc, $0.10$~pc), and ($0.25$~pc, $0.25$~pc) in vertical distance ranges of $|Z| \leq 1$~kpc, $1$~kpc $< |Z| \leq 2$~kpc, and $|Z| > 2$~kpc, respectively. The concentric dotted circles show heliocentric distances of $2$~kpc, $4$~kpc, and $6$~kpc. Right panel: same as in the left panel, but based on photometric metallicity estimates with $Gaia$ parallax priors.}
\label{fig:map} \end{figure*}

Figure~\ref{fig:map} shows an edge-on view of the metallicity distribution in the Galactocentric cylindrical coordinate system, where the Sun is located at ($R_\odot$, $Z_\odot$)=($8.34$~kpc, $20.8$~pc); \citep{reid:14,bennett:19}. The map in the left panel is based on the full photometric solution, while the right panel shows the metallicity map based on the $Gaia$ priors. I used an adapted mesh to explore the finer metallicity variations near the Galactic plane. Each pixel has a dimension ($\Delta R$, $\Delta Z$) of ($0.05$~pc, $0.05$~pc), ($0.10$~pc, $0.10$~pc), and ($0.25$~pc, $0.25$~pc) at $|Z| \leq 1$~kpc, $1$~kpc $< |Z| \leq 2$~kpc, and $|Z| > 2$~kpc, respectively. In each pixel of these maps, I estimated a weighted mean metallicity using a generalized histogram of a metallicity distribution function (MDF) with each star's metallicity weighted by its error; the resulting mean metallicity is robust against outliers. Iso-metallicity contours are overlaid to display detailed distributions.

The edge-on view of the disk in the left panel of Figure~\ref{fig:map} reveals a large-scale, wave-like distortion of the mean metallicity distribution in the radial direction. This behavior remains essentially unchanged even if I use extinctions higher by $10\%$ or restrict the sample to $E(B\, -\, V)<0.05$, although many stars at low Galactic latitudes are rejected by this selection. It is tempting to interpret this as a radial transverse density wave \citep{xu:15,schonrich:18} and/or a bending mode detected by {\it Gaia} \citep{gaiadr2_kinematics}, as would be expected from a passage of a Sagittarius dwarf galaxy in numerical simulations \citep[e.g.,][]{purcell:11,gomez:13}. However, because photometric zero-point errors, systematic errors in reddening, and systematic errors in extinction laws across the large survey area can produce spurious large-scale structures, the weak trend seen in Figure~\ref{fig:map} should be interpreted with caution.

\begin{figure*}
\centering
\includegraphics[scale=0.45]{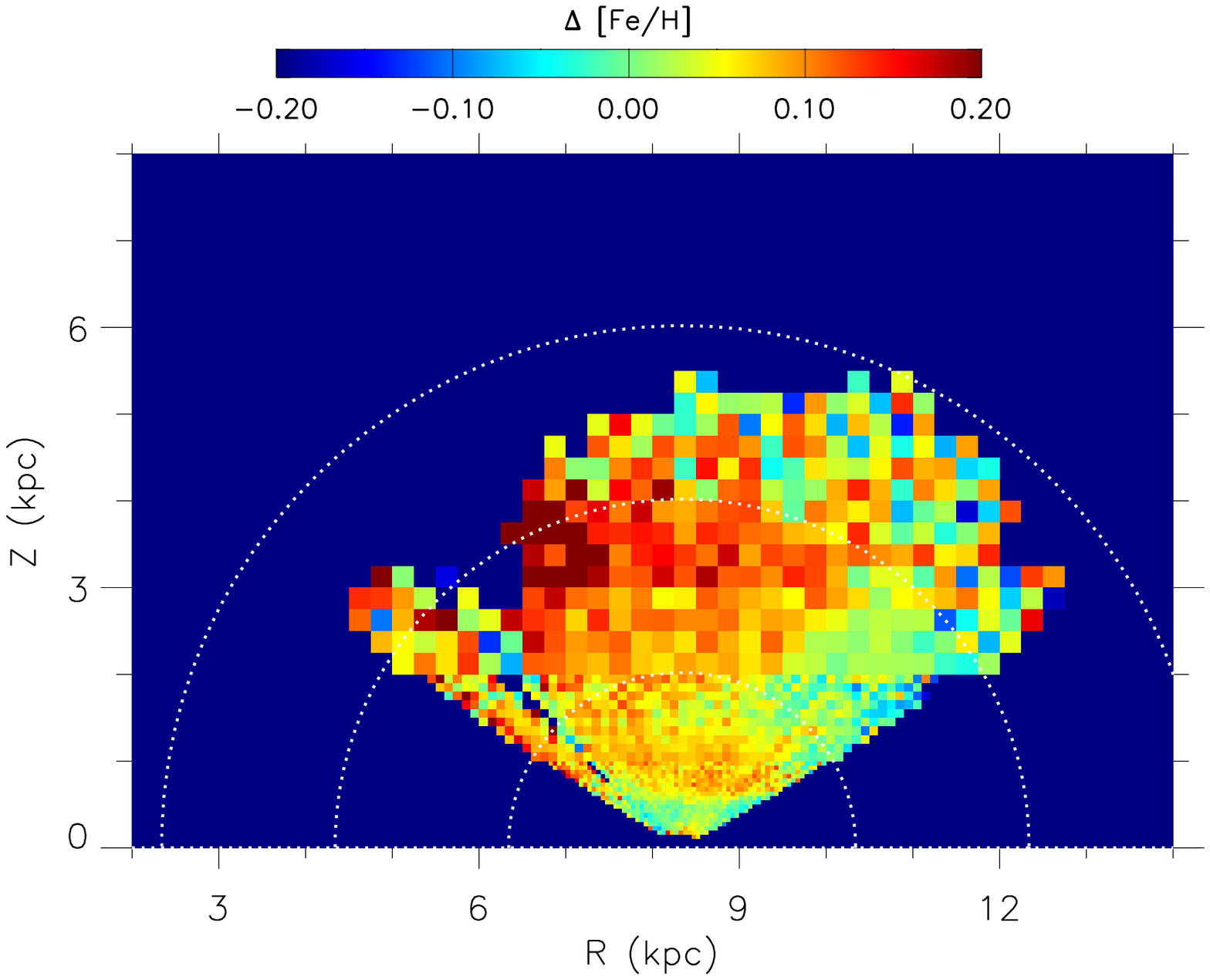}
\includegraphics[scale=0.45]{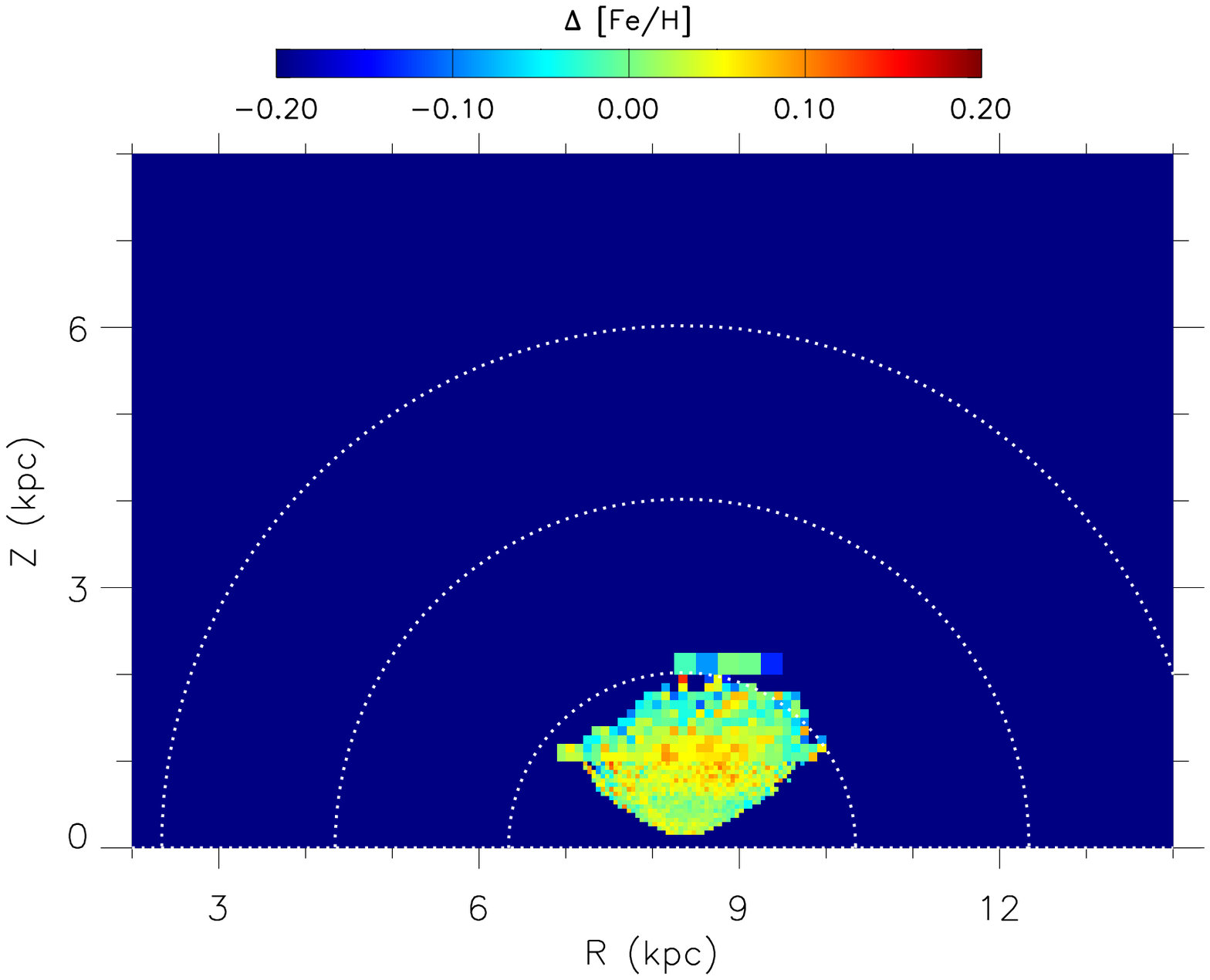}
\caption{Differences in the mean metallicities of stars between the northern and southern hemispheres, computed using the metallicity maps shown in Figure~\ref{fig:map}. The left panel is based on full photometric solutions, while the right panel is based on {\it Gaia} priors.}
\label{fig:diff} \end{figure*}

Figure~\ref{fig:diff} shows differences in the mean metallicities of stars between the northern and southern hemispheres (north minus south), computed using the metallicity maps shown in Figure~\ref{fig:map} at the same ($R$, $|Z|$). The left panel is based on the full photometric solutions, while the right panel is based on {\it Gaia} priors. The area with a positive $\Delta$[Fe/H] at $|Z|>2$~kpc depicts the tentative large-scale distortion as described above. In addition, Figure~\ref{fig:diff} shows a weaker oscillation pattern in the vertical direction, with a maximum difference seen at $|Z|\sim0.8$~kpc. Both photometric metallicity maps with and without {\it Gaia} priors exhibit the north--south asymmetry, with a similar amplitude at a similar vertical distance.

\begin{figure}
\centering
\includegraphics[scale=0.65]{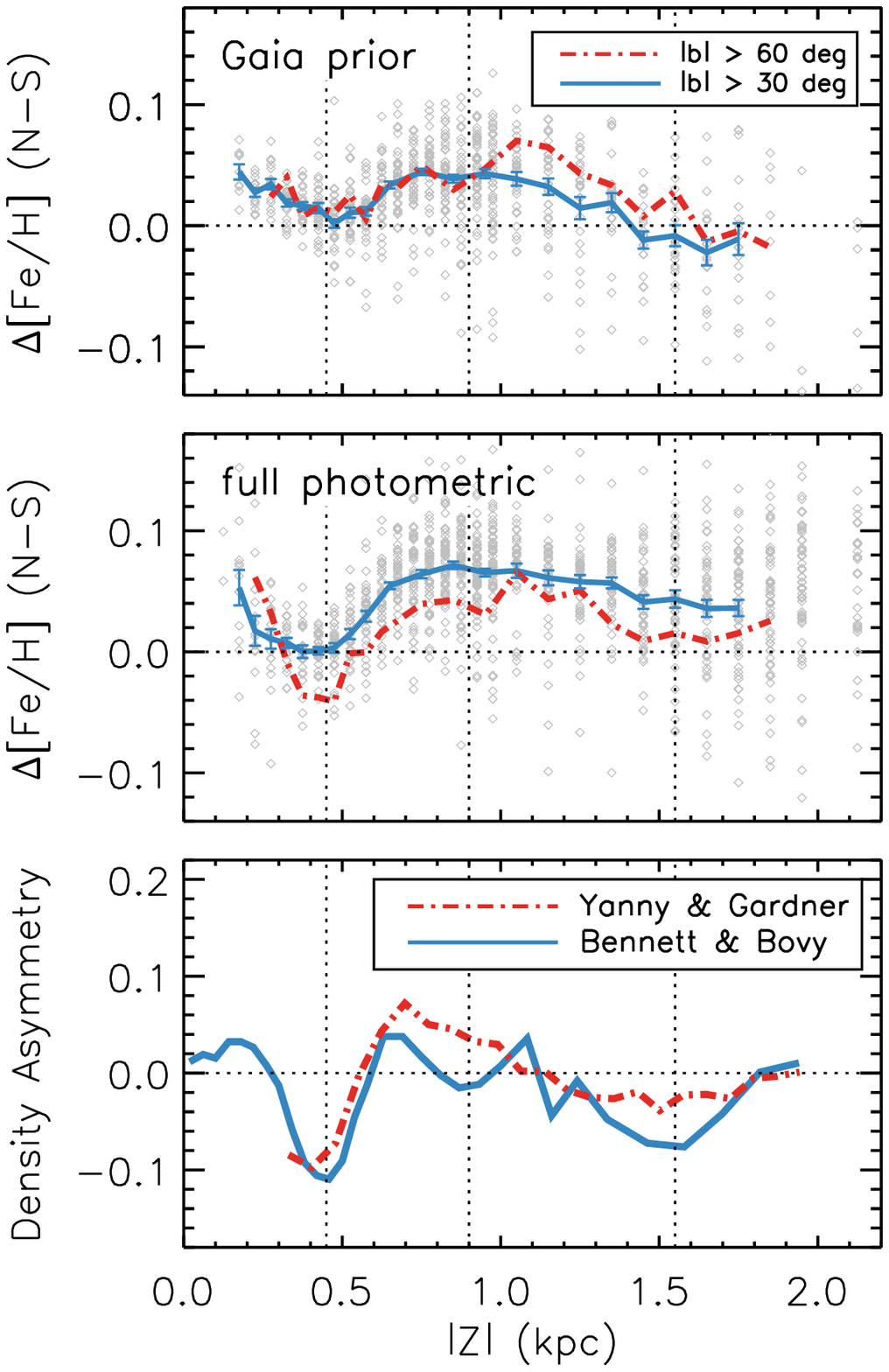}
\caption{Top panel: metallicity differences between the Galactic northern and southern hemispheres as a function of the vertical distance from the mid-plane (mean metallicities from north minus mean metallicities from south). Gray points indicate individual pixel values in the right panel in Figure~\ref{fig:diff}, showing differences in {\it Gaia}-based photometric metallicity of stars at $|b|>30\arcdeg$ with the same ($R$, $|Z|$) position in the Galactocentric cylindrical coordinate system. A solid blue curve shows moving averages of the differences, while the dotted--dashed red line shows a mean trend from stars at $|b|>60\arcdeg$. Middle panel: same as in the top panel, but based on a metallicity map from full photometric solutions without {\it Gaia} priors (left panel in Figure~\ref{fig:diff}). Bottom panel: vertical asymmetries of the stellar number density from \citet[][dotted--dashed red line]{yanny:13} and \citet[][solid blue line]{bennett:19}, respectively. Vertical dotted lines in all panels indicate the approximate locations of the three major troughs found in the latter study.}
\label{fig:asymmetry} \end{figure}

Metallicity differences projected onto $|Z|$ are shown in Figure~\ref{fig:asymmetry}. Gray points in the top panel indicate individual pixel values from {\it Gaia}-based photometric metallicities of stars (right panel in Figure~\ref{fig:diff}). The solid blue line shows moving averages in the difference using the same number of points in each moving box. The metallicity difference between the two hemispheres is nearly zero at $|Z|\approx450$~pc and increases to $\Delta{\rm [Fe/H]}=0.05$~dex at $|Z|\approx750$~pc. To check the effect of foreground extinction in the low Galactic latitude regions, I include the dotted--dashed red line that shows a mean trend observed from stars at $|b|>60\arcdeg$ (individual pixel values are not shown). The overall trend is similar to the case at $|b|>30\arcdeg$ except that there are small-scale structures when high-latitude stars are considered only. The middle panel shows the case without {\it Gaia} priors (left panel in Figure~\ref{fig:diff}). The overall trend remains qualitatively unchanged, even if I restrict the sample to stars with $E(B\, -\, V)<0.05$ or $10\%$ higher extinction values than in \citet{schlegel:98} are used.

There are some caveats in the photometric metallicity maps. The effect of photometric errors can be seen in Figure~\ref{fig:asymmetry} from a comparison between full photometric solutions and {\it Gaia}-based metallicity estimates. Full photometric solutions have larger errors in both distance and metallicity, which result in a steeper vertical metallicity gradient of the disk, and therefore a stronger metallicity asymmetry. In addition, the bright survey limit in the SDSS imaging data excludes a significant fraction of stars with $4.5 < M_r < 7.5$ close to the Galactic mid-plane. However, this limitation should be equally present in the data from both hemispheres, and its effect is likely canceled out in the metallicity difference. Finally, the zero-point difference in metallicity is not well defined due to small photometric zero-point differences between the northern and southern hemispheres \citep{schlafly:11},\footnote{The zero-point difference in $u$ is much more difficult to evaluate than in $griz$ because of the absence of independent, extensive $u$-band data \citep[see][]{finkbeiner:16}.} although the mean difference is only a few hundredth dex level in [Fe/H].

For comparison, the bottom panel in Figure~\ref{fig:asymmetry} shows differences in the number density of stars between the northern and southern Galactic hemispheres from \citet[][dotted--dashed red line]{yanny:13} and \citet[][solid blue line]{bennett:19}. The density asymmetry here is defined as $(N_{\rm North}-N_{\rm South})/(N_{\rm North}+N_{\rm South})$, where $N_{\rm North}$ and $N_{\rm South}$ represent a number density of stars in the northern and southern hemispheres, respectively. Many nearby stars were not included in \citet{yanny:13}, because stars brighter than $r\sim14$ were saturated in the SDSS imaging survey. Overall, the similar asymmetry patterns indicate large troughs at $|Z|\sim0.5$~kpc and $\sim1.5$~kpc, but the latter work reveals an additional trough at $|Z|\sim0.9$~kpc. These locations are marked by a vertical dotted line in all panels.

The trough in the density asymmetry at $|Z|\sim0.45$~kpc is coincident with the trough in the metallicity difference. At $|Z|\ga0.7$~kpc, the metallicity difference is slowly decreasing toward higher $|Z|$, which is consistent with the overall density asymmetry pattern. If I restrict the sample to high-latitude ($|b|>60\arcdeg$) stars, the small-scale troughs at $|Z|\sim0.9$~kpc and $|Z|\sim1.5$ also appear coincident with those in the density asymmetry. At $0.4$~kpc $< |Z| < 1.5$~kpc, the Pearson's correlation coefficients are $0.4 < r < 0.8$ between the density asymmetry in \citet{bennett:19} and the metallicity fluctuations for each subset of the sample shown in Figure~\ref{fig:asymmetry}.

\begin{figure*}
\centering
\includegraphics[scale=0.55]{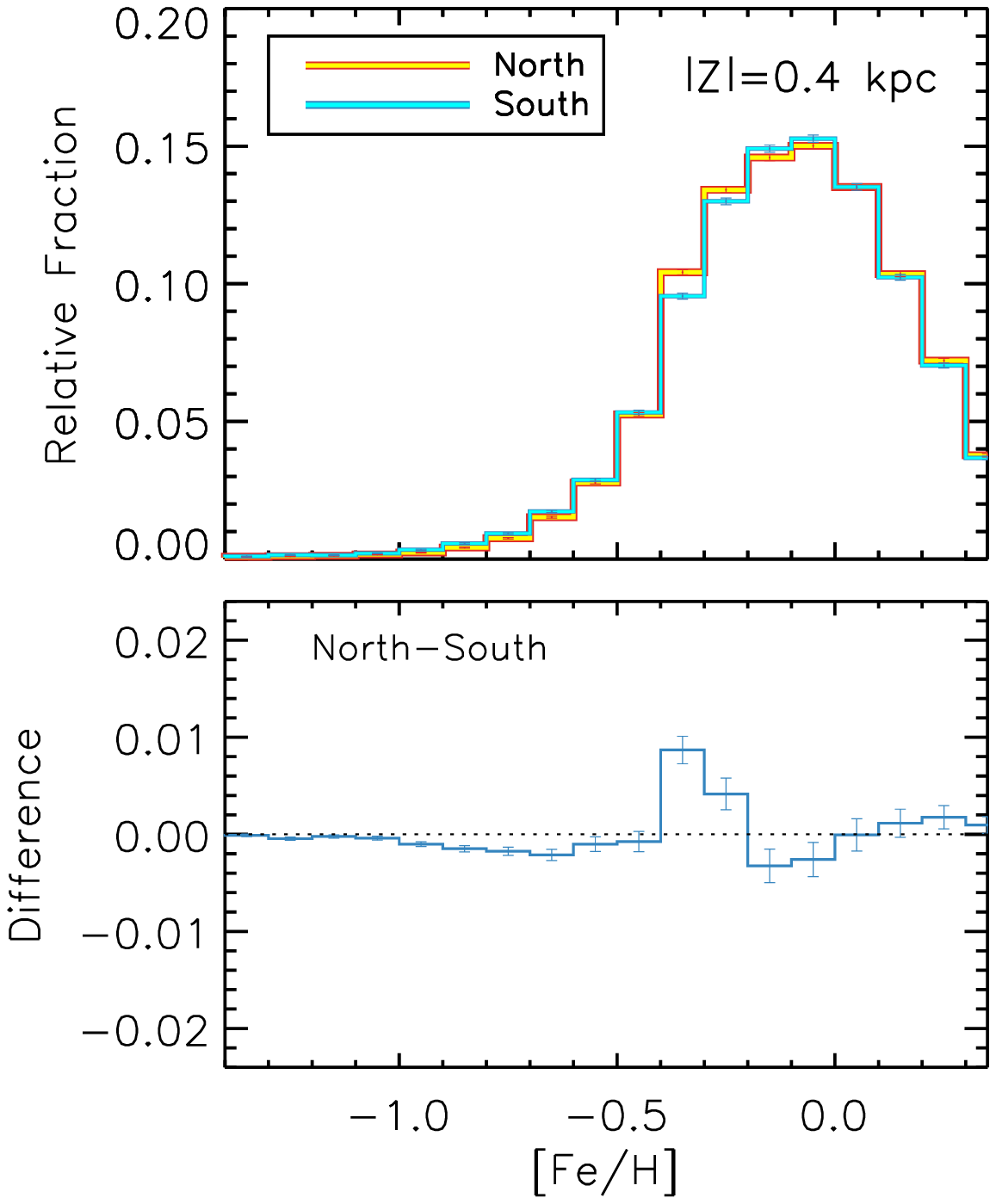}
\includegraphics[scale=0.55]{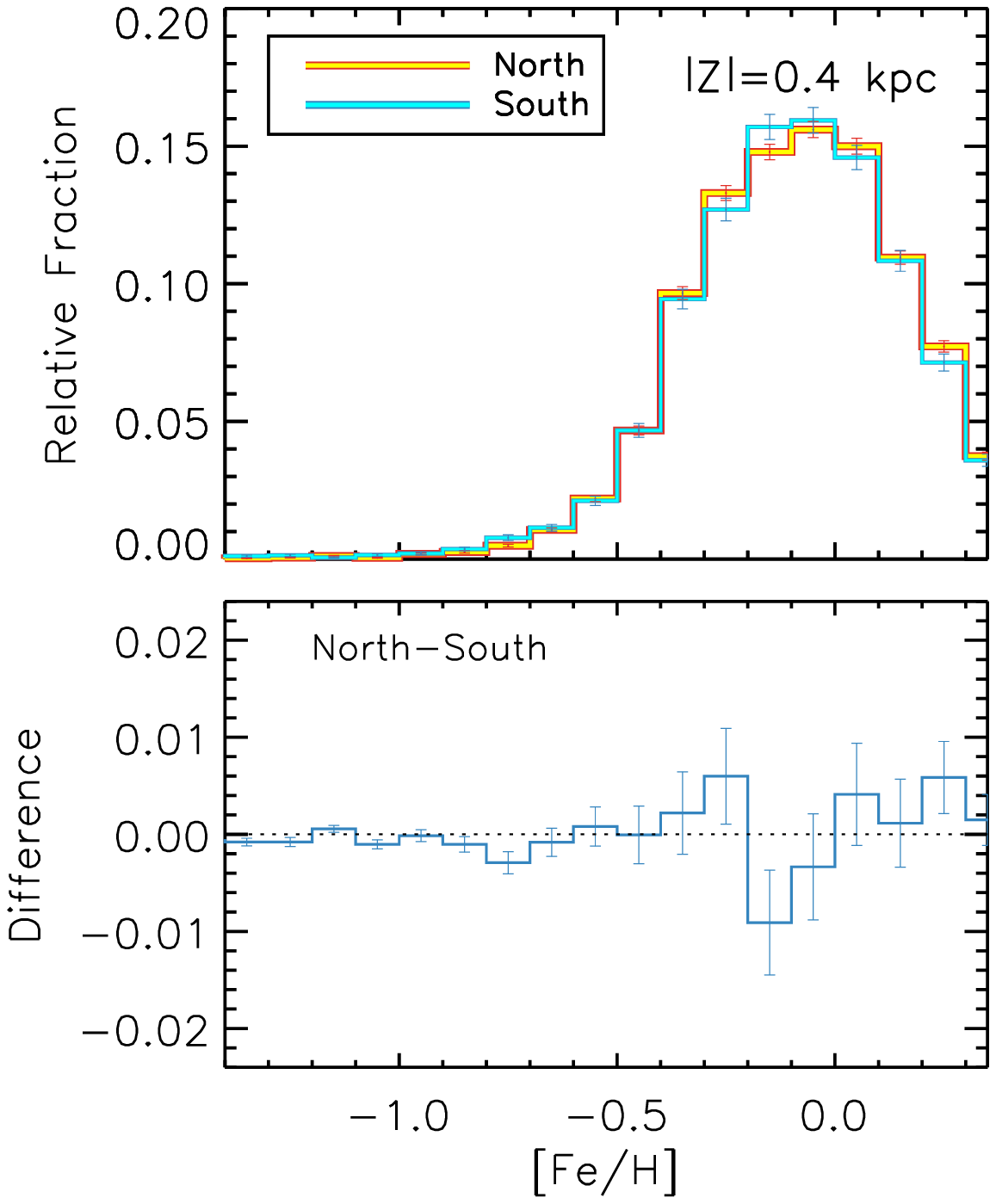}
\caption{Top panels: comparisons of the MDFs near the maximum trough in the vertical density asymmetry ($0.3$~kpc $< |Z| \leq 0.5$~kpc). The red and blue lines show the MDFs from the northern and southern Galactic hemispheres, respectively. The left panel is based on photometric metallicities with {\it Gaia} priors for stars at $|b|>30\arcdeg$, while the right panel shows the case at $|b|>60\arcdeg$. Bottom panels: fractional differences in the MDFs.}
\label{fig:mdf} \end{figure*}

The observed metallicity asymmetry is produced by a shift in the MDF as shown in Figure~\ref{fig:mdf}. The left and right panels show MDFs of stars near the first maximum trough ($0.3$~kpc $< |Z| \leq 0.5$~kpc) based on photometric metallicities with {\it Gaia} parallaxes at $|b|>30\arcdeg$ (left panel) and at $|b|>60\arcdeg$ (right panel), respectively. This region is dominated by thin disk stars (or $\alpha$-poor stars), and the observed MDF peaks near the solar metallicity. A close inspection reveals that the MDF from the northern hemisphere (red histogram) is shifted toward lower metallicities than the MDF in the southern hemisphere (blue histogram). The bottom panels show that these shifts can induce a small but noticeable difference in the mean metallicity of the sample.

\section{Summary and Discussion}

I present the first evidence of the metallicity asymmetry in the Galactic disk based on the metallicity distributions of stars obtained using the SDSS photometry. The present data displays the oscillatory behavior of the vertical metallicity structure, and shows a striking phase overlap with previously discovered density asymmetries between the northern and southern Galactic hemispheres. This result is robust against different sample selections and the use of {\it Gaia} parallaxes.

A simple plane wave assumption is inadequate to explain the observed phase overlap between the density and metallicity asymmetries. In the pressure-supported wave propagating through a disk in equilibrium, the displacement of stars is an even function at the maximum or minimum densities, and the net change in metallicity should become zero. Likewise, the density and metallicity asymmetries should be out of phase by a quarter wavelength, which is opposite to the finding presented in this work.

The phase overlap between the observed density and metallicity asymmetries can satisfactorily be explained by the phase-space wrapping of stars in the $Z$--$V_Z$ plane, which has recently been discovered by {\it Gaia} \citep{antoja:18,binney:18,blandhawthorn:19,darling:19,laporte:19}. As noted by these authors, the phase-space spiral is understood in terms of relaxation of disk stars from a bending perturbation, which has been excited by the tidal pull of the Milky Way's disk by a recent passage of a massive halo substructure like the Sagittarius dwarf galaxy \citep[but see][]{khoperskov:19}. Indeed, vertical distances with maximum/minimum density asymmetries approximately match turning points ($V_Z=0$) in the phase-space spiral \citep[$-590$~pc, $-230$~pc, $400$~pc, and $750$~pc;][]{antoja:18}, which give rise to the minimum density asymmetry at $|Z|\sim500$~pc and the maximum density asymmetry at $|Z|\sim700$~kpc. Because stars that constitute the phase-space spiral have been pulled away from the disk mid-plane, their mean metallicities become higher than those expected from the equilibrium disk. This can be seen in the comparison of MDFs (Figure~\ref{fig:mdf}), and is consistent with a fact that the phase-space spiral is more prominent for metal-rich stars \citep{blandhawthorn:19}. Therefore, it is expected that the density asymmetry should be positively correlated with the metallicity asymmetry, as I have found here.

\citet{antoja:18} presented the phase-space wrapping at $|Z|<1$~kpc. \citet{laporte:19} extended the volume of the phase-space spiral, tracing a ridge at $|Z|\sim1.1$~kpc and a trough at $|Z|\sim1.2$~kpc. This is seen as a density asymmetry in \citet{bennett:19}, although it is not clear in \citet{yanny:13}. The metallicity map presented in this work also reveals a small asymmetry in this distance range for stars at high Galactic latitudes (Figure~\ref{fig:asymmetry}). In addition, the density asymmetries suggest a dip at $|Z|\sim1.5$~kpc, which can be matched to the asymmetry in metallicity for the high-latitude stars from the full photometric solutions (see the middle panel in Figure~\ref{fig:asymmetry}). The feature is not well defined in the {\it Gaia}-based solution, because the number of stars with good parallax measurements is small at this distance range.

The maximum (peak-to-peak) amplitude at $|Z|\sim0.5$~kpc is $\Delta$[Fe/H]$\sim0.05$ from the {\it Gaia}-based solution. If the local disk has been displaced from the global mid-plane by a constant amount, the metallicity difference implies an offset in $Z$ of an order of $80$~pc, taking the mean metallicity gradient of the local disk stars ($\sim0.3$~dex~kpc$^{-1}$). This initial condition can be used to set constraints on the nature of the perturber and the local gravitational potential.

Characterizing the oscillations in a collisionless system is a complex problem due to the presence of the Galactic potential, mode damping, and a coupling with a nature of a perturber \citep[e.g.,][]{weinberg:91,widrow:14,blandhawthorn:19,darling:19,laporte:19}. Understanding the detailed mixing of disk stars and the evolution of the Galactic disk will greatly benefit from the observational data presented in this work and will also benefit from future photometric surveys like the Large Synoptic Survey Telescope \citep{ivezic:19}.

\acknowledgements

I thank Timothy Beers for his invaluable comments on the photometric metallicity mapping of the Milky Way. I acknowledge support provided by Basic Science Research Program through the National Research Foundation of Korea (NRF) funded by the Ministry of Education (NRF-2018R1D1A1A02085433) and by the Korean NRF to the Center for Galaxy Evolution Research (No.\ 2017R1A5A1070354). I acknowledge the use of SDSS data (https://www.sdss.org/).

{}

\end{document}